\documentclass[aps,prl,showpacs,showkeys]{revtex4-1}

\usepackage{amsmath,amssymb,mathrsfs,bbm}
\usepackage{graphicx}
\usepackage{amsthm}
\usepackage{amscd}
\usepackage{bm}
\usepackage{dsfont}
\usepackage[utf8]{inputenc}

\usepackage{wasysym}

\begin{document}

\title{Superfluid neutron matter \\in the $s$-channel exchange nucleon-nucleon interaction models
%\footnote{This work was presented as
%a part of invited talks at
%Conference on Isospin, Structure, Reactions and Energy of Symmetry (ISTROS), Častá - Papiernička, Častá, Slovakia, May 14 - 19, 2017
%and Workshop on Spectroscopy and Reactions of Exotic Nuclei (Huzhou-CUSTIPEN), Huzhou, China, July 3 - 9, 2017.}
}
\author{M. I. Krivoruchenko$^{1,2,3}$}
%EndAName
\affiliation{
$^{1}$Institute for Theoretical and Experimental Physics$\mathrm{,}$ B. Cheremushkinskaya 25\\
117218 Moscow, Russia}
\affiliation{
$^{2}$Moscow Institute of Physics and Technology$\mathrm{,}$ 141700 Dolgoprudny$\mathrm{,}$ Russia}
\affiliation{
$^{3}$Bogoliubov Laboratory of Theoretical Physics$\mathrm{,}$ Joint Institute for Nuclear Research\\
141980 Dubna$\mathrm{,}$ Russia
}
%\date{}

%\begin{center}
%{\bf \large Abstract}
%\end{center}

\begin{abstract}
The superfluid pairing gap of neutron matter is calculated in the framework of
Quark Compound Bag model with nucleon-nucleon interactions generated by the $s$-channel exchange of Jaffe-Low primitives (6-quark states).
\end{abstract}

%\keywords{nucleon-nucleon interaction, nuclear matter, neutron pairing, QCB model}

\maketitle

%\newpage

%\baselineskip= 13pt

%%%%%%%%%%%%%%%%%%%%%%%%%%%%%%%%%%%%%%%%%%%%%%%%%%%%%%%%%%%%%%%%%%%%%%%%%%%%%%%%%%%%%%%%%%%%%%%%%%%%%%%%%%%%%%%%%%%%%%%%%%%%%%%%%%%%%%%%%%
\section{I. Introduction}
%\renewcommand{\theequation}{I.\arabic{equation}}
%\setcounter{equation}{0}
%%%%%%%%%%%%%%%%%%%%%%%%%%%%%%%%%%%%%%%%%%%%%%%%%%%%%%%%%%%%%%%%%%%%%%%%%%%%%%%%%%%%%%%%%%%%%%%%%%%%%%%%%%%%%%%%%%%%%%%%%%%%%%%%%%%%%%%%%%

It is well known that nucleon-nucleon interactions are characterized by repulsion at small distances and attraction at large distances. In the one-boson exchange (OBE) model, the main role is attributed to the $\omega $-meson exchange, which generates repulsion, and the $\sigma $-meson exchange, which generates attraction.
There are many versions of the OBE model with a different set of mesons that describe successfully a broad set of experimental data (see, e.g., \cite {1,1a,2,3}). The strong interaction scale, however, is comparable to the size of nucleons and mesons. From a geometric point of view, it is not quite clear how the $t$-channel exchange mechanism of the OBE model can dominate in those cases where the overlap of particles becomes essential. Simple estimates show that, in nuclei, the overlap of nucleons is rather substantial \cite {4}.

In situations where an overlap of nucleons is important, the quark degrees of freedom should be taken into account.
Two nucleons at small distances form a 6-quark state. The interaction of nucleons, therefore, can be described by a diagram
where nucleons propagate first, and then merge together in a 6-quark state; this state propagates and then decays into two nucleons (see, e.g., Fig.~\ref{fig2}).

This mechanism was initially discussed by T.~D.~Lee \cite{6} without connection to the nucleon-nucleon scattering problem.
In order to illustrate the physical nature of Castillejo, Dalitz and Dyson poles \cite{8}, Dyson \cite{7} constructed a modified Lee model with the $s$-channel exchange of resonances.
Dyson-Lee models allow for the existence of bound states and resonances, so they describe a class of systems dominated by attraction, whereas the short-range nucleon-nucleon interactions are dominated by repulsion. Simonov \cite{9} expanded the class of Dyson-Lee models by including the $s$-channel exchange of Jaffe-Low "primitives" \cite{10}. The extended Simonov-Dyson models describe systems with both attraction and repulsion, and the repulsion is generated by Jaffe-Low primitives. The primitives can be interpreted as resonances with a vanishing width on the mass shell. Off the mass shell, the primitive widths are different from zero, so these states are involved in the interactions. Primitives correspond to zeros of the scattering phase with a negative slope, while the $P$ matrix has poles at the energy of nucleons equal to the primitive mass. At the same time, the $S$ matrix is a regular function.

In a class of Simonov-Dyson models, the most detailed studies are made in the framework of Quark Compound Bag (QCB) model. This model successfully reproduces the nucleon-nucleon scattering phases, as well as the properties of light nuclei \cite{9,9a,9b,9c,11,12,13,14}.

In this paper, we calculate the superfluid pairing gap of neutron matter in one of the versions of QCB model \cite{13}.

The neutron matter superfluidity is of interest for
modeling glitches of neutron stars,
describing cooling rate and the structure of neutron stars.
The equation of state (EoS) of nuclear matter is of interest for astrophysics of compact objects. The discovery of neutron stars with a mass of about $2 M_{\astrosun}$ \cite{15,16} allowed exclusion of a broad set of the soft EoS of nuclear matter for which neutron stars lose gravitational stability at lower values of the masses.
At the same time, laboratory experiments indicate that the EoS of the symmetric nuclear matter must be soft enough.
This problem has been widely discussed in recent years. An important role in the EoS of nuclear matter is attributed
to the isospin asymmetry, which influences the stiffness with the increase in neutron fraction \cite{Li:2014oda}.
The production of hyperons in the centre of massive neutron stars due to the chemical equilibrium with respect to the weak interactions
softens the EoS \cite{17,18}. The possibility of increasing the stiffness of the EoS of nuclear matter
at the expense of introducing weakly interacting light bosons (WILBs) beyond the standard model has been discussed
in Refs. \cite{19,20,21}. In Ref. \cite{20}, we pointed out an additional source of repulsion between hyperons,
associated with the $\phi  $(1020)-meson exchange, which is normally suppressed in interactions of non-strange baryons due
to the Okubo-Zweig-Iizuka rule. The numerical studies of Refs.~\cite{23,24} verify that the $\phi$(1020)-meson
suppresses the hyperon production efficiently enough to keep the maximum neutron star mass above the observational limit.
The laboratory data still leave a certain freedom for stiff high-density EoS of nuclear matter in the chemical equilibrium.

In the mean field (MF) approximation, the exotic degrees of freedom soften the EoS of nuclear matter.
The upper limit on the masses of neutron stars $2 M_{\astrosun}$ leads to strong restrictions on the critical
density of the phase transition to quark matter and, if quark matter exists in the cores of neutron stars,
to the quark matter EoS. Beyond the MF approximation, the effect of exotic degrees of freedom is, in general,
multidirectional. In quantum theory, even if critical transition density into a new phase is high,
exotic degrees of freedom are present virtually and contribute to observables through loops.
This requires their account already at the saturation density and leads to a renormalization of the phenomenological parameters.
When the density increases, the sign of the effect is not fixed \textit{a priori}.
An example is given in Ref. \cite{25}, where a dibaryon Bose condensation in nuclear matter is discussed
in the relativistic Hartree approximation \cite{26}. In this regard, one can expect
that the effect of recently discovered dibaryon $d^{*}$(2380) \cite{27,28,29,30,31,32} on the EoS of nuclear matter,
despite its relatively high mass, is important because the spin of the resonance, $J = 3$, provides a large $2J + 1$-fold enhancement of the Casimir effect
originating from the in-medium modification of zero-point fluctuations of the dibaryon field.
A dibaryon Bose condensation in nuclear matter is discussed in Refs. \cite{14,25,33,34,35,36,37,38,39,40,41,42,43,44,Aguirre:1998vh,Vidana:2017qey}.

The problem of how nuclear matter behaves with increasing the density in the $s$-channel exchange models of nucleon-nucleon interaction
has not yet been studied. In the present paper, we formulate and solve the equations for the neutron matter EoS
in the framework of QCB model of Ref. \cite{13} and determine the density dependence of the neutron pairing gap.
The model parameters of Ref.~\cite{13} are fitted to
the nucleon-nucleon scattering phases for the nucleon kinetic energy in the laboratory frame
up to 350 MeV in the $^1 S_0$ channel and up to 500 MeV in the $^3 S_1$ channel.
In such an approach, the quantitative description of nuclear matter
at a density below the saturation density must be possible.
Moreover, we believe that extrapolation of the model predictions to supranuclear densities
gives at least a qualitatively correct picture of nuclear matter phenomena.

This paper is organized as follows. In the next section, the model is described.
Section 3 provides a system of equations to determine the EoS by taking into account the neutron pairing effect.
Section 4 describes a procedure for numerical solution of the equations and reports results of the calculations.
In Conclusion, the results are summarized and perspectives of studying
high-density nuclear matter in the $s$-channel exchange nucleon-nucleon interaction models
are discussed.

%%%%%%%%%%%%%%%%%%%%%%%%%%%%%%%%%%%%%%%%%%%%%%%%%%%%%%%%%%%%%%%%%%%%%%%%%%%%
%%%%%%%%%%%%%%%%%%%%%%%%%%%%%%%%%%%%%%%%%%%%%%%%%%%%%%%%%%%%%%%%%%%%%%%%%%%%
%%%%%%%%%%%%%%%%%%%%%%%%%%%%%%%%%%%%%%%%%%%%%%%%%%%%%%%%%%%%%%%%%%%%%%%%%%%%
\begin{figure} [t] %
\begin{center}
\includegraphics[angle = 270,width=0.62\textwidth]{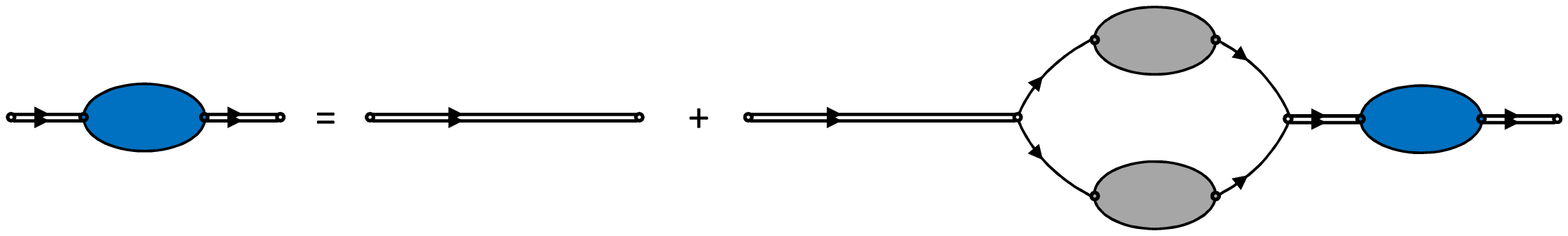}
\caption{(color online)
Dyson equation for the compound-state Green's function.
The bare Green's function $1/\Lambda(s)$ is represented by a double line.
The complete Green's function  $1/D(s)$ is represented by a double line with a blue-shaded block.
The complete Green's function of the nucleons is represented by a solid line with a grey-shaded block.
}
\label{fig1}
\end{center}
\end{figure}
%\vspace{-10mm}
%%%%%%%%%%%%%%%%%%%%%%%%%%%%%%%%%%%%%%%%%%%%%%%%%%%%%%%%%%%%%%%%%%%%%%%%%%%%
%%%%%%%%%%%%%%%%%%%%%%%%%%%%%%%%%%%%%%%%%%%%%%%%%%%%%%%%%%%%%%%%%%%%%%%%%%%%
%%%%%%%%%%%%%%%%%%%%%%%%%%%%%%%%%%%%%%%%%%%%%%%%%%%%%%%%%%%%%%%%%%%%%%%%%%%%

%%%%%%%%%%%%%%%%%%%%%%%%%%%%%%%%%%%%%%%%%%%%%%%%%%%%%%%%%%%%%%%%%%%%%%%%%%%%%%%%%%%%%%%%%%%%%%%%%%%%%%%%%%%%%%%%%%%%%%%%%%%%%%%%%%%%%%%%%%
%%%%%%%%%%%%%%%%%%%%%%%%%%%%%%%%%%%%%%%%%%%%%%%%%%%%%%%%%%%%%%%%%%%%%
\section{II. Model}
%\renewcommand{\theequation}{II.\arabic{equation}}
%\setcounter{equation}{0}
%%%%%%%%%%%%%%%%%%%%%%%%%%%%%%%%%%%%%%%%%%%%%%%%%%%%%%%%%%%%%%%%%%%%%%%%%%%%%%%%%%%%%%%%%%%%%%%%%%%%%%%%%%%%%%%%%%%%%%%%%%%%%%%%%%%%%%%%%%
%%%%%%%%%%%%%%%%%%%%%%%%%%%%%%%%%%%%%%%%%%%%%%%%%%%%%%%%%%%%%%%%%%%%%

In the QCB model \cite{9,9a,9b,9c,11,12,13,14}, the nucleon-nucleon scattering proceeds through the formation of an intermediate
compound state with a mass of  $M_{\alpha} > \sqrt{s_0} = 2m$, where $m$ is the nucleon mass. The $D$ function of the process can be written as
\begin{equation}
D(s) = \Lambda(s) - \Pi(s),
\label{1}
\end{equation}
where
\begin{eqnarray}
\Lambda^{-1}(s) &=& \sum_{\alpha} \frac{g_{\alpha}^2}{s - M_{\alpha}^2} + G, \nonumber \\
\Pi(s)     &=& - \frac{1}{\pi}\int_{s_0}^{+\infty}\Phi_{2}(s^{\prime})\frac{\mathcal{F}^2(s^{\prime})}{s^{\prime} - s}ds^{\prime}, \label{2}
\end{eqnarray}
Here $\Phi_{2}(s) = \pi p^{*}/\sqrt{s}$ is the relativistic phase space, $p^{*} = \sqrt{s/4 - m^2}$  is the nucleon momentum in the center-of-mass frame,  $g_{\alpha}$ is the coupling constant of primitive $d_{\alpha}$ with nucleons, $\mathcal{F}(s)$  is form factor of the $d_{\alpha}NN$ vertex. The $S$ matrix has the form
\begin{equation}
S = \frac{D(s - i0)}{D(s + i0)}. \label{3}
\end{equation}
The poles of $\Lambda(s)$, known as Castillejo, Dalitz and Dyson poles \cite{10}, are localized between zeros of the function $\Lambda(s)$, which, respectively,
are determined by the masses of compound states from equation $\Lambda(s = M_{\alpha}^2) = 0$. After the coupling with the continuum is
switched on ($g_{\alpha} \neq 0$), compound states become bound states, resonances, or primitives \cite{7,13}.

The $D$ function (\ref{1}) is a generalized $R$ function. It does not have complex zeros on the first sheet of the Riemann surface.
On the real half-axis $(-\infty,s_0)$, when the condition $D(s_0) < 0$ is satisfied for $s_0 < M_{\alpha}^2$, it also does not have zeros
describing bound states.

Simple roots of equation
\begin{equation}
D(s) = 0, \label{4}
\end{equation}
located under the unitary cut on an nonphysical sheet of the Riemann surface, are identified with resonances.
Simple roots of the equation, located on the real half-axis $(s_0,+\infty)$, are identified with primitives.
In such a case, the zeros of the real part of $D(s)$ are of the first order, and the zeros of the imaginary part of $D(s)$ are of the second order.

The value $1/D(s)$ can be interpreted as a complete propagator of the compound state (or states).
The complete propagator can be determined from the Dyson equation as shown in Fig.~\ref{fig1}.
The loop in the diagram denotes the dispersive part of the $D$ function, i.e., $\Pi(s)$. In a more general scheme,
there is a contact four-fermion interaction term, to which the coupling constant $G$ in Eq. (\ref{2}) corresponds.

The vertex $d_{\alpha}NN$ corresponds to the value $-ig_{\alpha}\mathcal{F}(s)$, the bare compound-state propagator
is in the correspondence with $i/(s - M_{\alpha}^2) +iG/g_{\alpha}^2$,
and the complete propagator is $1/D(s)$. The contact vertex is included in the definition of the bare propagator.
The form factor  $\mathcal{F}(s)$ is a function of the three-dimensional momentum of nucleons in the center-of-mass frame,
while the compound-state propagator depends on their four-momenta through $s = (p_1 + p_2)^2$.
On the mass shell of nucleons $\mathcal{F}(s)$ is the function of nucleon momentum: $\mathcal{F}(s = 4(m^2 + \mathbf{p}^2)) \equiv \mathcal{F}(\mathbf{p}^{2} )$.

%%%%%%%%%%%%%%%%%%%%%%%%%%%%%%%%%%%%%%%%%%%%%%%%%%%%%%%%%%%%%%%%%%%%%%%%%%%%
%%%%%%%%%%%%%%%%%%%%%%%%%%%%%%%%%%%%%%%%%%%%%%%%%%%%%%%%%%%%%%%%%%%%%%%%%%%%
%%%%%%%%%%%%%%%%%%%%%%%%%%%%%%%%%%%%%%%%%%%%%%%%%%%%%%%%%%%%%%%%%%%%%%%%%%%%
\begin{figure} [t] %
\begin{center}
\includegraphics[angle = 0,width=0.538\textwidth]{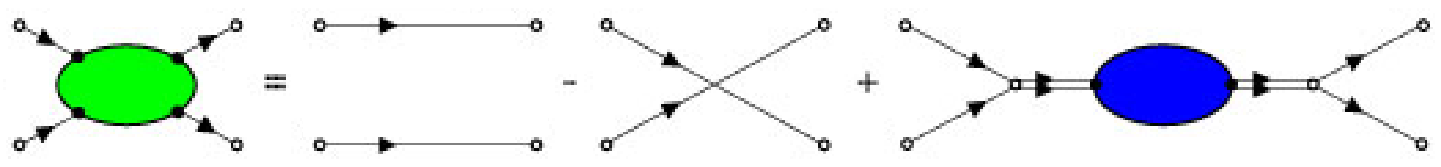}
\caption{(color online)
Graphical representation of two-nucleon Green's function.
The compound-state Green's function $1/D(s)$
determines the $T$ matrix.
}
\label{fig2}
\end{center}
\end{figure}
%\vspace{-10mm}
%%%%%%%%%%%%%%%%%%%%%%%%%%%%%%%%%%%%%%%%%%%%%%%%%%%%%%%%%%%%%%%%%%%%%%%%%%%%
%%%%%%%%%%%%%%%%%%%%%%%%%%%%%%%%%%%%%%%%%%%%%%%%%%%%%%%%%%%%%%%%%%%%%%%%%%%%
%%%%%%%%%%%%%%%%%%%%%%%%%%%%%%%%%%%%%%%%%%%%%%%%%%%%%%%%%%%%%%%%%%%%%%%%%%%%

The scattering of two nucleons is shown in Fig.~\ref{fig2}. The amplitude has the form
\begin{eqnarray}
A(s) &=& e^{i\delta(s)} \sin \delta(s) \nonumber \\
     &=& - \frac{\Phi_{2}(s) \mathcal{F}^2(s)}{D(s)}. \label{5}
\end{eqnarray}
In the Born approximation, we have the relation
\begin{eqnarray}
A(s) &=& - \frac{\sqrt{s}p^{*}}{8\pi}U(\mathbf{q}) \nonumber \\
     &=& - \Phi_{2}(s) \mathcal{F}(s) \Lambda^{-1}(s) \mathcal{F}(s). \label{6}
\end{eqnarray}
where $\mathbf{q} = \mathbf{p}^{\prime} - \mathbf{p}$ is the transmitted momentum, and $U(\mathbf{q})$ is the $s$-wave projected Fourier transform of the potential.
The scattering theory in separable potentials can be used;
the kinematic factors are restored from the correspondence
\begin{equation}
U(\mathbf{q}) \leftrightarrow \frac{8\pi^2}{s}\mathcal{F}(s) \Lambda^{-1}(s) \mathcal{F}(s). \label{7}
\end{equation}
Separable potentials are represented in the form  $U(\mathbf{q}) = \sum_{\nu} f_{\nu}(\mathbf{p}^{\prime})f_{\nu}(\mathbf{p})$.

%%%%%%%%%%%%%%%%%%%%%%%%%%%%%%%%%%%%%%%%%%%%%%%%%%%%%%%%%%%%%%%%%%%%%%%%%%%%%%%%%%%%%%%%%%%%%%%%%%%%%%%%%%%%%%%%%%%%%%%%%%%%%%%%%%%%%%%%%%
\section{III. Neutron pairing}
%\renewcommand{\theequation}{III.\arabic{equation}}
%\setcounter{equation}{0}
%%%%%%%%%%%%%%%%%%%%%%%%%%%%%%%%%%%%%%%%%%%%%%%%%%%%%%%%%%%%%%%%%%%%%%%%%%%%%%%%%%%%%%%%%%%%%%%%%%%%%%%%%%%%%%%%%%%%%%%%%%%%%%%%%%%%%%%%%%

For separable potentials, the pairing gap equations are discussed in Ref. \cite{45} and recently in Ref. \cite{46}.
In view of the correspondence (\ref{7}), the self-consistency condition can be written as
\begin{equation}
1=-\int  \frac{d\mathbf{p}}{(2\pi )^{3} } \frac{2\pi^{2} }{E^{2}(\mathbf{p})} \mathcal{F}(\mathbf{p}^{2}) \Lambda^{-1}(s)\mathcal{F}(\mathbf{p}^{2})
   \left. \frac{1}{2\sqrt{(E(\mathbf{p}) - \mu )^{2} +\Delta^{2} (s,\mathbf{p})} } \right|_{s = 4 \mu^{2} } .     \label{8}
\end{equation}
Here $s$ is the (square) of the energy of the Cooper pair in its rest frame. The value of $s$ is set equal to 4$\mu^2 $, where $\mu $ is the chemical
potential of neutrons; the relativistic dispersion law $E(\mathbf{p})=\sqrt{\mathbf{p}^{2} +m^{2} } $ is used. The pairing gap equals
\begin{equation}
\Delta (4\mu ^{2} ,\mathbf{p}) = \frac{\sqrt{2} \pi }{E(\mathbf{p})} \mathcal{F}(\mathbf{p}^{2} )\Lambda ^{-1}(4\mu ^{2} )|\Xi |.
\label{9}
\end{equation}
The value $\Xi ^{*} $ is defined by
\begin{equation} \label{10}
i\Xi ^{*} =\int \frac{d^{4} p}{(2\pi )^{4} }  \frac{\sqrt{2} \pi }{E(\mathbf{p})} \mathcal{F}(\mathbf{p}^{2} )F^{\dag } (p),
\end{equation}
with $F^{\dag } (p)$ being the anomalous Green's function in the momentum-space (see, e.g., \cite{47}):
\begin{eqnarray*}
\varepsilon_{\alpha \beta } F^{\dag } (p) &=& \int d^{4}x e^{ip(x - y)} \varepsilon_{\alpha \beta }F^{\dag } (x-y) \nonumber \\
                                          &=& \int d^{4}x e^{ip(x - y)} (-i)\left\langle T\Psi _{\alpha }^{\dag } (x)\Psi _{\beta }^{\dag } (y) \right\rangle,
\end{eqnarray*}
where $\varepsilon _{12} = - \varepsilon _{21} =1$, and $\varepsilon_{11} =\varepsilon _{22} =0.$
Equations (\ref{8}) - (\ref{10}) can be derived from Eliashberg's equations for normal and anomalous Green's functions \cite{47,50}.
In Fig.~\ref{fig20} Eliashberg's equations are shown graphically for the $s$-channel exchange interaction models.

The solution of the self-consistency equation leads to the following expression for the normal Green's function
\begin{equation}
G(p) = \frac{u_{\mathbf{p}}^{2} }{\omega -\varepsilon (\mathbf{p}) + i0} + \frac{v_{\mathbf{p}}^{2} }{\omega + \varepsilon (\mathbf{p}) - i0}, \label{11}
\end{equation}
where
\begin{equation*}
\left(\begin{array}{c} {u_{\mathbf{p}}^{2} } \\ {v_{\mathbf{p}}^{2} } \end{array}\right) = \frac{1}{2} \left(1 \pm \frac{\eta _{\mathbf{p}} }{\varepsilon (\mathbf{p})} \right),
\end{equation*}
and $\eta _{\mathbf{p}} = E(\mathbf{p}) - \mu$, $\varepsilon (\mathbf{p}) = \sqrt{\eta_{\mathbf{p}}^{2}  + \Delta^{2}(4\mu ^{2} ,\mathbf{p})}$. The anomalous Green's function equals
\begin{equation} \label{12}
F^{\dag } (p) = -\frac{\sqrt{2} \pi }{E(\mathbf{p})} {\mathcal F}(\mathbf{p}^{2} ) \Lambda^{-1} (4\mu ^{2} ) \frac{\Xi^{*} }{(\omega -\varepsilon (\mathbf{p}) + i0)(\omega + \varepsilon(\mathbf{p}) - i0)} .
\end{equation}
Substituting Eq.~(\ref{12}) into Eq.~(\ref{10}), we arrive at (\ref{8}).

%%%%%%%%%%%%%%%%%%%%%%%%%%%%%%%%%%%%%%%%%%%%%%%%%%%%%%%%%%%%%%%%%%%%%%%%%%%%
%%%%%%%%%%%%%%%%%%%%%%%%%%%%%%%%%%%%%%%%%%%%%%%%%%%%%%%%%%%%%%%%%%%%%%%%%%%%
%%%%%%%%%%%%%%%%%%%%%%%%%%%%%%%%%%%%%%%%%%%%%%%%%%%%%%%%%%%%%%%%%%%%%%%%%%%%
\begin{figure} [t] %
\begin{center}
\includegraphics[angle = 0,width=0.108\textwidth]{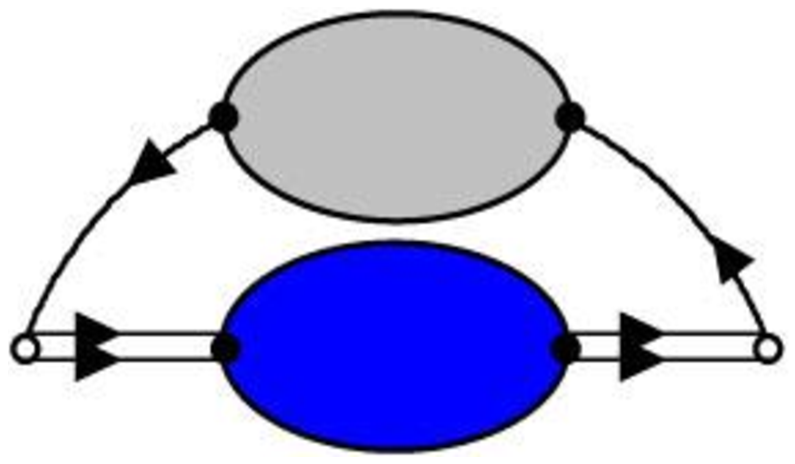}
\caption{(color online)
The nucleon proper self-energy part, or mass operator $\Sigma$ in nuclear matter.
The loop is formed by a hole in the Fermi sphere and the compound state.
}
\label{fig3}
\end{center}
\end{figure}
\vspace{2mm}
%%%%%%%%%%%%%%%%%%%%%%%%%%%%%%%%%%%%%%%%%%%%%%%%%%%%%%%%%%%%%%%%%%%%%%%%%%%%
%%%%%%%%%%%%%%%%%%%%%%%%%%%%%%%%%%%%%%%%%%%%%%%%%%%%%%%%%%%%%%%%%%%%%%%%%%%%
%%%%%%%%%%%%%%%%%%%%%%%%%%%%%%%%%%%%%%%%%%%%%%%%%%%%%%%%%%%%%%%%%%%%%%%%%%%%

Given that Green's function (\ref{11}) is known, the number density can be found from
\begin{eqnarray}
\frac{N}{V} &=& -2i \mathop{\lim }\limits_{t \to -0} \int  \frac{d\omega d\mathbf{p}}{(2\pi )^{4} } e^{-i\omega t} G(\omega ,\mathbf{p}) \nonumber \\
            &=& 2\int  \frac{d\mathbf{p}}{(2\pi )^{3} } v_{\mathbf{p}}^{2} = -\frac{\partial \Omega }{\partial \mu } . \label{13}
\end{eqnarray}
After integrating this equation with respect to the chemical potential, thermodynamic potential $\Omega $ can be found, then the energy of the system can be calculated from
\begin{equation} \label{14}
E = \Omega +\mu N.
\end{equation}

The primitive contributes to the self-energy of the in-medium nucleon, as shown in Fig.~\ref{fig3}. The contribution has
the structure (1 + $\gamma_0 $)/2, where $\gamma_0 $ is the Dirac gamma matrix \cite{14}. The mass
operator $\Sigma$ redefines the nucleon energy,

%%%%%%%%%%%%%%%%%%%%%%%%%%%%%%%%%%%%%%%%%%%%%%%%%%%%%%%%%%%%%%%%%%%%%%%%%%%%
%%%%%%%%%%%%%%%%%%%%%%%%%%%%%%%%%%%%%%%%%%%%%%%%%%%%%%%%%%%%%%%%%%%%%%%%%%%%
%%%%%%%%%%%%%%%%%%%%%%%%%%%%%%%%%%%%%%%%%%%%%%%%%%%%%%%%%%%%%%%%%%%%%%%%%%%%
\vspace{10mm}
\begin{figure} [h] %
\begin{center}
\includegraphics[angle = 270,width=0.62\textwidth]{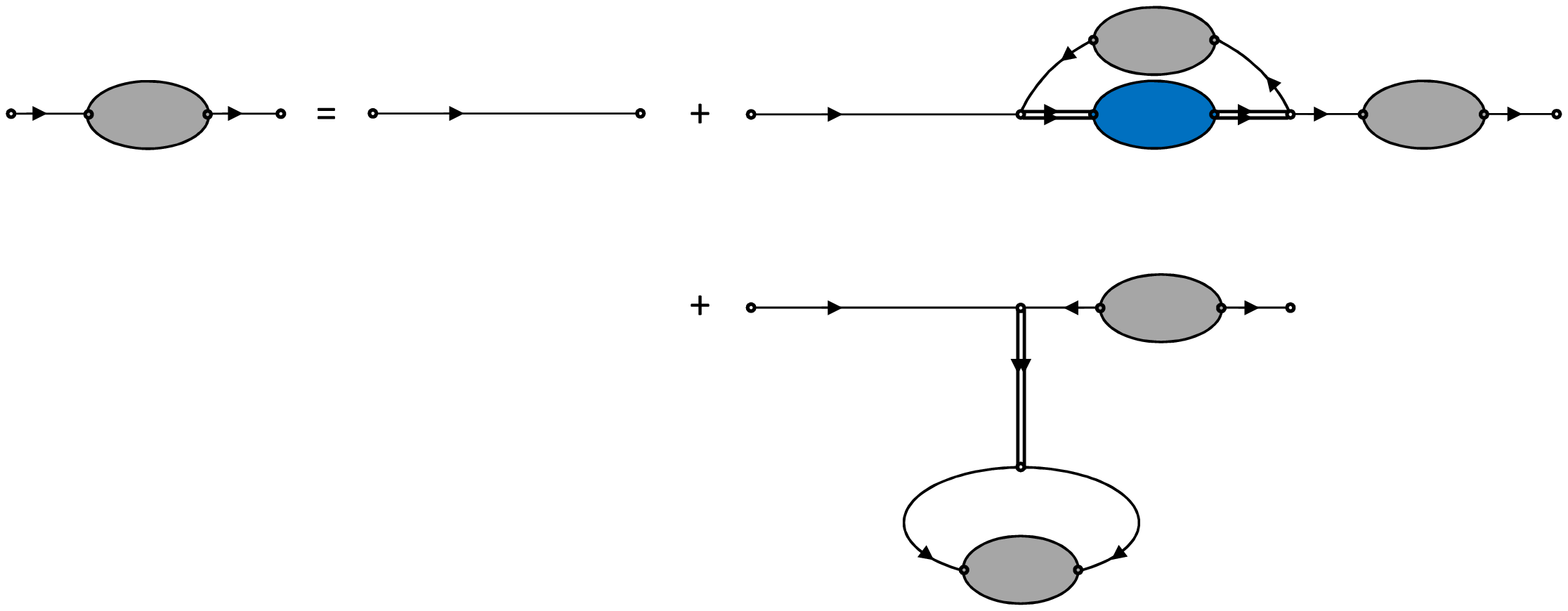}
\end{center}
\end{figure}
\vspace{5mm}
%%%%%%%%%%%%%%%%%%%%%%%%%%%%%%%%%%%%%%%%%%%%%%%%%%%%%%%%%%%%%%%%%%%%%%%%%%%%
%%%%%%%%%%%%%%%%%%%%%%%%%%%%%%%%%%%%%%%%%%%%%%%%%%%%%%%%%%%%%%%%%%%%%%%%%%%%
%%%%%%%%%%%%%%%%%%%%%%%%%%%%%%%%%%%%%%%%%%%%%%%%%%%%%%%%%%%%%%%%%%%%%%%%%%%%
%%%%%%%%%%%%%%%%%%%%%%%%%%%%%%%%%%%%%%%%%%%%%%%%%%%%%%%%%%%%%%%%%%%%%%%%%%%%
%%%%%%%%%%%%%%%%%%%%%%%%%%%%%%%%%%%%%%%%%%%%%%%%%%%%%%%%%%%%%%%%%%%%%%%%%%%%
%%%%%%%%%%%%%%%%%%%%%%%%%%%%%%%%%%%%%%%%%%%%%%%%%%%%%%%%%%%%%%%%%%%%%%%%%%%%
\begin{figure} [h] %
\vspace{5mm}
\begin{center}
\includegraphics[angle = 270,width=0.62\textwidth]{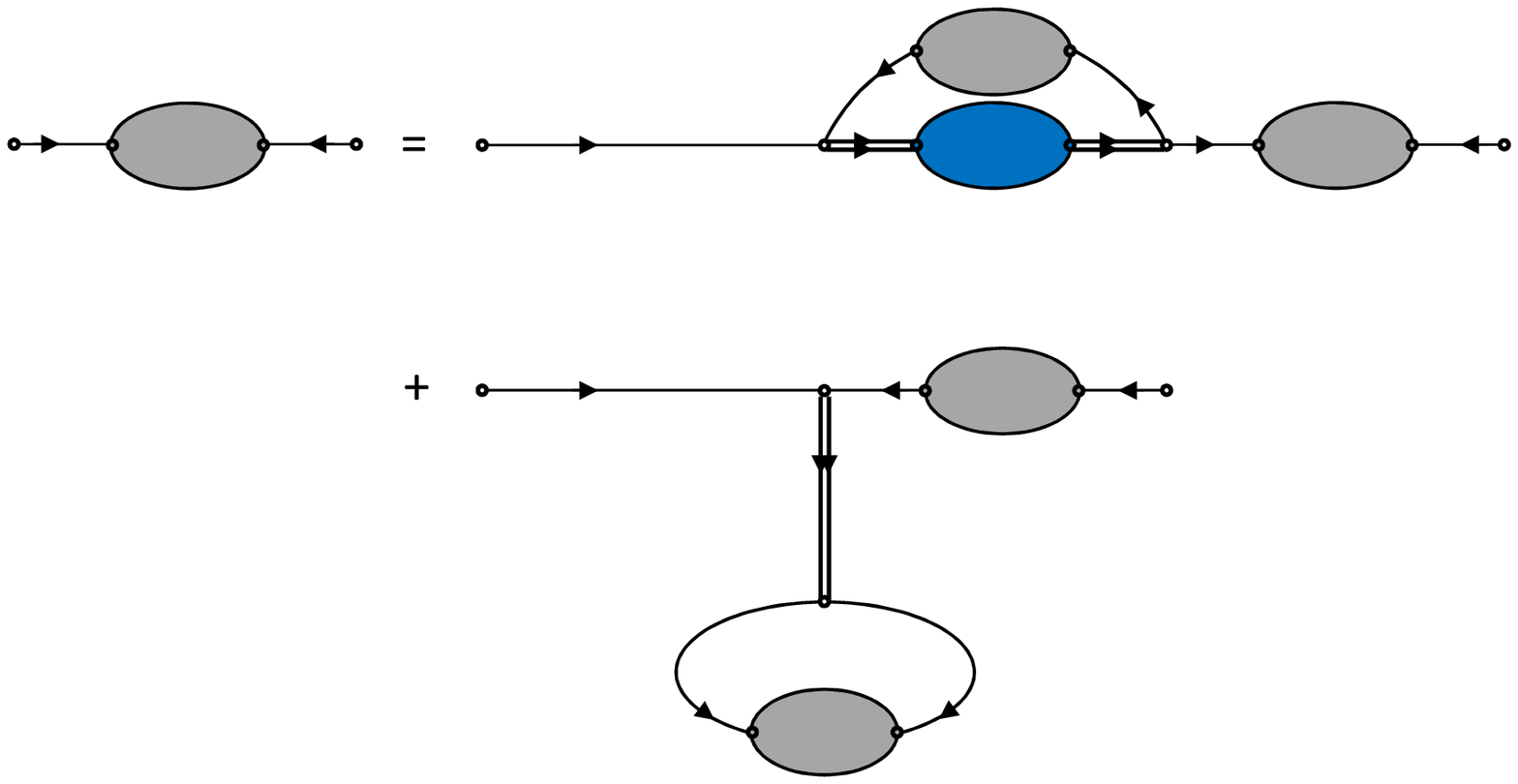}
\caption{(color online)
Graphical representation of Eliashberg's equations for normal and anomalous Green's functions
in the $ s $-channel exchange interaction models. The bare nucleon Green's function is drawn with a solid line.
The complete Green's function of the nucleons $ G (p) $ is represented by a solid line with a gray-shaded block and two arrows pointing in the same direction.
The anomalous Green's functions of nucleons are represented by single lines with gray-shaded blocks and two
incoming arrows for $ F^{\dag} (p) $ or two outgoing arrows for $ F (p) $. A closed loop formed by the anomalous Green's function $ F^{\dag} (p) $
determines the value of $ \Xi^{*} $ in the equation~(\ref{10}). The compound-state Green's functions are represented by double lines as in Figs.~\ref{1} - \ref{3}.
}
\label{fig20}
\end{center}
\end{figure}
%\vspace{-10mm}
%%%%%%%%%%%%%%%%%%%%%%%%%%%%%%%%%%%%%%%%%%%%%%%%%%%%%%%%%%%%%%%%%%%%%%%%%%%%
%%%%%%%%%%%%%%%%%%%%%%%%%%%%%%%%%%%%%%%%%%%%%%%%%%%%%%%%%%%%%%%%%%%%%%%%%%%%
%%%%%%%%%%%%%%%%%%%%%%%%%%%%%%%%%%%%%%%%%%%%%%%%%%%%%%%%%%%%%%%%%%%%%%%%%%%%
$\;$\\
i.e., the chemical potential, and contributes to the nucleon mass. The dispersion law is taken to be
\begin{equation} \label{15}
E(\mathbf{p})=\Sigma/2 + \sqrt{\mathbf{p}^{2} + (m + \Sigma/2)^{2}} .
\end{equation}

The mass operator has therefore vectorial and scalar components in the Lorentz group. In order to discriminate them from each other,
it is necessary to go beyond the non-relativistic approximation, as discussed in Ref. \cite{14}. The mass operator, in general,
depends on the nucleon momentum and is defined off the energy surface,
which makes the dispersion law a more complicated function in comparison with (\ref{15}). In the mass operator, we neglect
the shift from the energy surface, as well as the momentum dependence. When calculating $\Sigma$, the nucleon momentum is set to zero. Figure ~\ref{fig3} corresponds to the expression
\begin{equation} \label{16}
\Sigma =\frac{2 \pi^{2} }{m^{2} } \int \frac{d\mathbf{p}}{(2\pi )^{3} }  \frac{{\mathcal F}^{2} (s)}{D(s)} v_{\mathbf{p}}^{2} ,
\end{equation}
where the function $v_{\mathbf{p}}^{2} $  is the probability of finding a nucleon with a given momentum. Equation (\ref{16}) generalizes the corresponding equation of the optical potential model. Here, $\mathbf{p}$ is the momentum of the nucleon in the rest frame of the matter, $s = (m + E(\mathbf{p}))^2 - \mathbf{p}^2$
is the square of the nucleons energy in the rest frame of the matter.
The in-medium modification of the $ T $ matrix is not considered, namely, when calculating the loop in Fig.~\ref{1},
the Pauli blocking for nucleons is not taken into account, and the imaginary part of the nucleon self-energy is discarded.

In a free theory, the chemical potential is determined by the Fermi momentum from
\begin{equation*}
p_{F}^{[0]} =\sqrt{\mu ^{2} -m^{2} } .
\end{equation*}
Using condition (\ref{15}) one finds a momentum at which the quasiparticle energy is minimal. The equation $E(\textbf{p}) = \mu$ gives
\begin{equation*}
p_{F}^{[1]} =\sqrt{(\mu -\Sigma /2)^{2} -(m+\Sigma /2)^{2} } .
\end{equation*}
If the density is known, the Fermi momentum can also be found from equation
\begin{equation} \label{17}
n=\frac{2}{(2\pi )^{3} } \frac{4\pi }{3} p_{F}^{3} .
\end{equation}
In a theory with interaction, these three momenta are pairwise different. The minimum of $\varepsilon(\textbf{p})$ determines
the energy gap of the quasiparticle spectrum:
\begin{equation*}
\Delta _{F} \equiv \Delta (4\mu ^{2} ,p_{F}^{[1]} ).
\end{equation*}

%%%%%%%%%%%%%%%%%%%%%%%%%%%%%%%%%%%%%%%%%%%%%%%%%%%%%%%%%%%%%%%%%%%%%
\section{IV. Numerical results}
%\renewcommand{\theequation}{IV.\arabic{equation}}
%\setcounter{equation}{0}
%%%%%%%%%%%%%%%%%%%%%%%%%%%%%%%%%%%%%%%%%%%%%%%%%%%%%%%%%%%%%%%%%%%%%

The calculation scheme is as follows: We vary the chemical potential and look for solutions of $\Xi ^{*} $ in equation (\ref{8}). As the starting value of $\Sigma $, we calculate the integral (\ref{16}) for $v_{\mathbf {p}}^{2} = 1$ inside and $v_{\mathbf {p}}^{2} = 0$ outside the Fermi sphere. For a given $\Sigma $, we find $\Xi ^{*} $. Next, we find $v_{\mathbf {p}}^{2} $, calculate the self-energy $\Sigma $, find $\Xi ^{*} $, and so on until the convergence becomes obvious.

The particle density can be found according to Eq.~(\ref{13}). Further, by integrating the density with respect to the chemical potential, one can find the thermodynamic potential, which is the pressure with the opposite sign. In conclusion, according to Eq.~(\ref{14}), the energy of the system can be calculated.

For the model parameters of Ref. \cite{13}, the results of solving Eq.~(\ref{8}) are shown in Fig.~\ref{fig5},
where $\Delta_F $ is a function of the Fermi momentum. Predictions of the QCB model are superimposed on the predictions of the advanced OBE models \cite{48}.

%%%%%%%%%%%%%%%%%%%%%%%%%%%%%%%%%%%%%%%%%%%%%%%%%%%%%%%%%%%%%%%%%%%%%%%%%%%%
%%%%%%%%%%%%%%%%%%%%%%%%%%%%%%%%%%%%%%%%%%%%%%%%%%%%%%%%%%%%%%%%%%%%%%%%%%%%
%%%%%%%%%%%%%%%%%%%%%%%%%%%%%%%%%%%%%%%%%%%%%%%%%%%%%%%%%%%%%%%%%%%%%%%%%%%%
%\vspace{ 5mm}
\begin{figure} [h] %
\begin{center}
\includegraphics[angle = -90,width=0.38\textwidth]{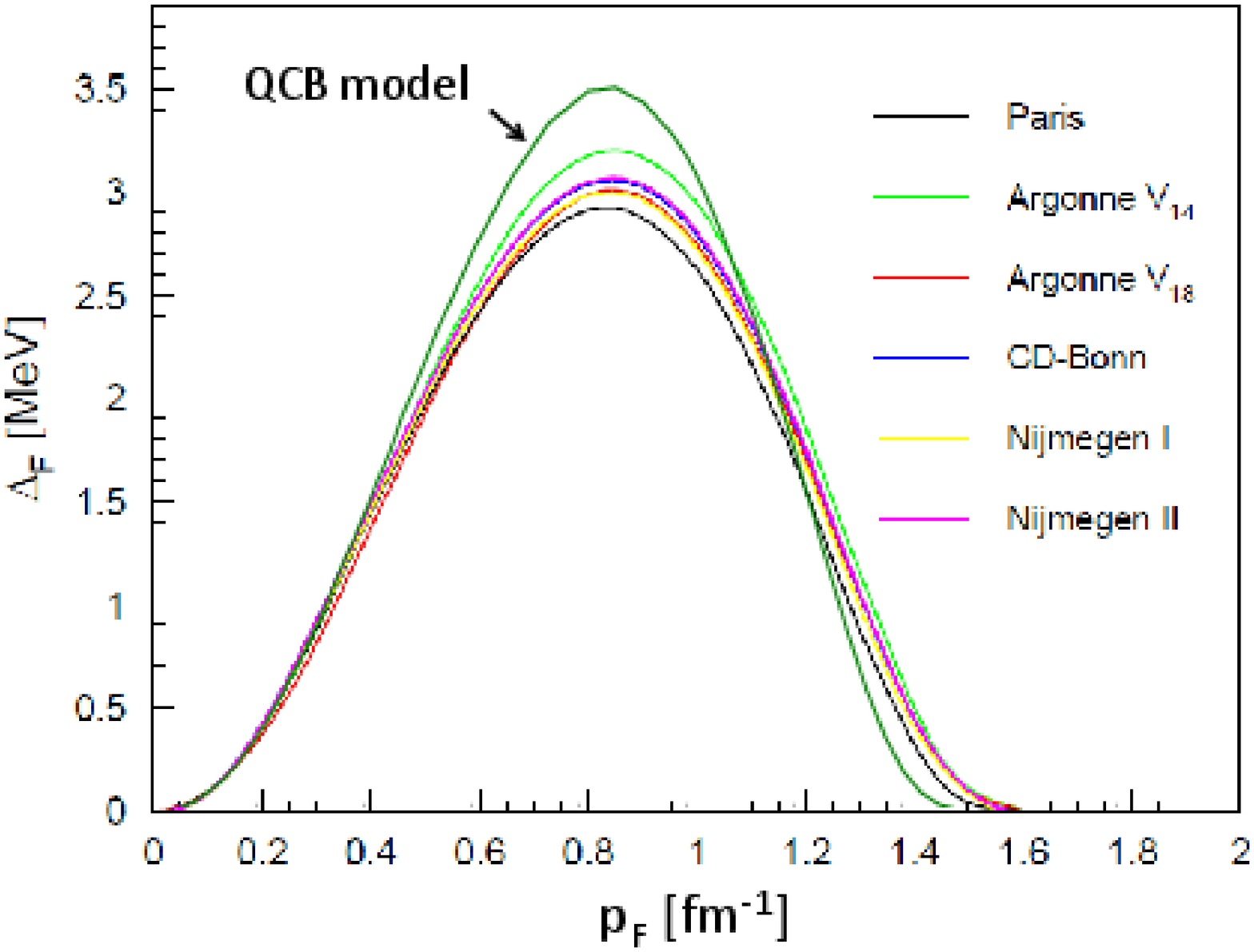}
\caption{
The neutron pairing gap $\Delta_F $ versus the Fermi momentum in neutron matter. The QCB model predictions are compared with the results of OBE models for free meson spectral functions: Paris, Argonne V$_{14}$, Argonne V$_{18}$, CD-Bonn, Nijmegen I, and Nijmegen II \cite{48}.
}
\label{fig4}
\end{center}
\end{figure}
%\vspace{-10mm}
%%%%%%%%%%%%%%%%%%%%%%%%%%%%%%%%%%%%%%%%%%%%%%%%%%%%%%%%%%%%%%%%%%%%%%%%%%%%
%%%%%%%%%%%%%%%%%%%%%%%%%%%%%%%%%%%%%%%%%%%%%%%%%%%%%%%%%%%%%%%%%%%%%%%%%%%%
%%%%%%%%%%%%%%%%%%%%%%%%%%%%%%%%%%%%%%%%%%%%%%%%%%%%%%%%%%%%%%%%%%%%%%%%%%%%

In the model considered, the vanishing of the energy gap at a momentum $p_F > 1.6$ fm$^{-1}$ is related to the zero of the form factor ${\mathcal F}(s)$ at $p^* = 353$ MeV. When the momentum $p_{F}^{[1]} $ approaches $p^*$, the smallness in the denominator of the integrand (\ref{8}) is compensated by the smallness of the numerator, as a result of which the integral remains small and the solutions do not appear.

With a further increase in the chemical potential, $p_{F}^{[1]} $ shifts relative to the zero of ${\mathcal F}(s)$ at $p^* = 353$ MeV, which results in the occurrence
of an additional branch of solutions at high densities, as shown on a different scale in Fig.~\ref{fig5}.
The negative sign of the pairing gap is due to the fact that the form factor changes sign (The pairing gap as a function of the momentum is proportional to the form factor, see Eq. (\ref{9}). In this region, the dibaryon chemical potential is still below 2007 MeV, so the solutions have the usual physical meaning.

%%%%%%%%%%%%%%%%%%%%%%%%%%%%%%%%%%%%%%%%%%%%%%%%%%%%%%%%%%%%%%%%%%%%%%%%%%%%
%%%%%%%%%%%%%%%%%%%%%%%%%%%%%%%%%%%%%%%%%%%%%%%%%%%%%%%%%%%%%%%%%%%%%%%%%%%%
%%%%%%%%%%%%%%%%%%%%%%%%%%%%%%%%%%%%%%%%%%%%%%%%%%%%%%%%%%%%%%%%%%%%%%%%%%%%
%\vspace{-5mm}
\begin{figure} [h] %
\begin{center}
\includegraphics[angle = 0,width=0.38\textwidth]{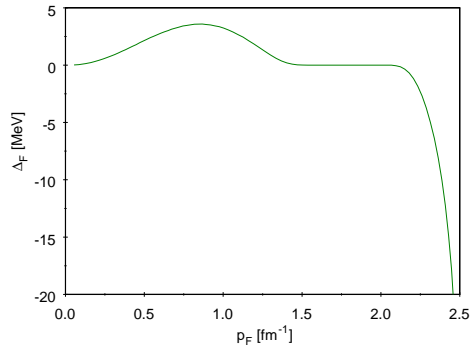}
\caption{
The superfluid pairing gap of neutron matter versus the Fermi momentum in a reduced scale. For $p_F > 2.0$ fm$^{-1}$,
a new branch of solutions of Eq.~(\ref{8}) occurs.
The value of $p_F = 2.5$ fm$^{-1}$ corresponds to the mass of the primitive 2007 MeV.}
\label{fig5}
\end{center}
\end{figure}
%%%%%%%%%%%%%%%%%%%%%%%%%%%%%%%%%%%%%%%%%%%%%%%%%%%%%%%%%%%%%%%%%%%%%%%%%%%%
%%%%%%%%%%%%%%%%%%%%%%%%%%%%%%%%%%%%%%%%%%%%%%%%%%%%%%%%%%%%%%%%%%%%%%%%%%%%
%%%%%%%%%%%%%%%%%%%%%%%%%%%%%%%%%%%%%%%%%%%%%%%%%%%%%%%%%%%%%%%%%%%%%%%%%%%%

\begin{figure} [h] %
\begin{center}
\includegraphics[angle = 0,width=0.38\textwidth]{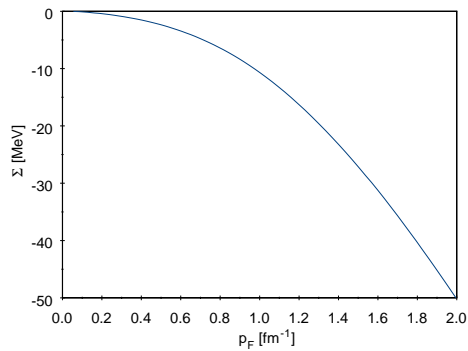}
\caption{
The neutron self-energy as a function of the Fermi momentum.
}
\label{fig6}
\end{center}
\end{figure}

The effective interaction constant is proportional to $\Lambda^{-1}(4\mu ^{2} )$, so one can expect that the energy gap increases with $\mu $. This effect together with the existence of a new branch of solutions are clearly seen in Fig.~\ref{fig6} for $p_F> 2.0$ fm$^{-1}$.
This behavior has a nice interpretation in terms of repulsion in a two-level system.
The first level is the Cooper pair, and the second higher level is the primitive (compound state) with the same quantum numbers. The interaction between these states leads to the repulsion, and the effect does not depend on the sign of the potential. Since the primitive is not treated dynamically, its mass is fixed. The binding energy of the Cooper pair, consequently, increases with $\mu $, which leads to an increase in the pairing gap.

When the Fermi momentum is less than 1.6 fm$^{-1}$, the energy gap is consistent with the OBE models. Small densities correspond to large distances, which are well studied and well parameterized in all models .

On the horizontal axis, in Figures~\ref{fig4} - \ref{fig6}, a Fermi momentum is shown, which is determined from the particle number density (\ref{17}). The region of applicability of the model is limited to momenta $p < p_F = 2.5$ fm$^{-1}$. At a Fermi momentum of the order of 2.5 fm$^{-1}$, a dibaryon Bose condensation can start due to the primitive-to-resonance conversion.

%%%%%%%%%%%%%%%%%%%%%%%%%%%%%%%%%%%%%%%%%%%%%%%%%%%%%%%%%%%%%%%%%%%%%%%%%%%%
%%%%%%%%%%%%%%%%%%%%%%%%%%%%%%%%%%%%%%%%%%%%%%%%%%%%%%%%%%%%%%%%%%%%%%%%%%%%
%%%%%%%%%%%%%%%%%%%%%%%%%%%%%%%%%%%%%%%%%%%%%%%%%%%%%%%%%%%%%%%%%%%%%%%%%%%%
%\vspace{-5mm}
%\vspace{-10mm}
%%%%%%%%%%%%%%%%%%%%%%%%%%%%%%%%%%%%%%%%%%%%%%%%%%%%%%%%%%%%%%%%%%%%%%%%%%%%
%%%%%%%%%%%%%%%%%%%%%%%%%%%%%%%%%%%%%%%%%%%%%%%%%%%%%%%%%%%%%%%%%%%%%%%%%%%%
%%%%%%%%%%%%%%%%%%%%%%%%%%%%%%%%%%%%%%%%%%%%%%%%%%%%%%%%%%%%%%%%%%%%%%%%%%%%

Figure~\ref{fig6} shows the dependence of the neutron self-energy operator on the Fermi momentum.
The modification of the neutron mass is much smaller than predicted in the MF models, and is comparable in order
of magnitude with the predictions of the DBHF models.

More details on the nuclear matter properties in the model considered can be found elsewhere \cite{52a}.

%%%%%%%%%%%%%%%%%%%%%%%%%%%%%%%%%%%%%%%%%%%%%%%%%%%%%%%%%%%%%%%%%%%%%
\section{V. Conclusions}
%%%%%%%%%%%%%%%%%%%%%%%%%%%%%%%%%%%%%%%%%%%%%%%%%%%%%%%%%%%%%%%%%%%%%

In this paper, we studied the superfluid neutron matter in the $s$-channel exchange nucleon-nucleon interaction model of Ref.~\cite{13}.
The neutron pairing gap below the saturation density was found to be fairly consistent with previous studies in the OBE models.

The results obtained by the extrapolation to supranuclear densities have transparent physical interpretations,
including the origin of the second superfluid phase of the neutron substance. In the model considered,
a dibaryon Bose condensation is possible at a density of 0.55 fm$^{-3}$, i.e., three times greater than the saturation density.

The presented calculations can be improved in several ways:

1.~In Refs. \cite{9,9a,9b,9c,11,12,14}, the scattering phases of nucleons were described in a broad energy interval by introducing two primitives in each channel.
Instead of one self-consistency equation (\ref{8}), here two self-consistency equations must be considered.
We expect that the qualitative properties of nuclear matter will not be changed, but EoS may become stiffer.
Next, we discussed one $^1 S_0$ scattering channel of neutrons. In a more advanced approach, one should include other channels.

2.~Evaluation of properties of the symmetric nuclear matter at the saturation density is a necessary step in studying the nuclear matter EoS. Properties of the symmetric nuclear matter are experimentally known, so this calculation is a sensitive test of the model.

3.~In the present paper, the primitive was not treated dynamically, so that the
the region of applicability of the model is restricted to densities $n < 0.55$ fm$^{-3}$.
In the absence of special constraints and under perturbations primitives, in general, leave the unitary cut and become resonances,
which leads to a Bose condensation of 6-quark resonances, i.e., dibaryons.
In OBE models, primitives are tightly coupled to the unitary cut.
The same physics in QCB models could require either fine tuning of the model parameters
or a special constraint.
The instability of primitives under perturbations allows for experimental verification \cite{49}.

4.~In this paper we found an approximate solution of superfluidity equations of neutron matter.
The pairing gap was calculated in terms of the $T$ matrix. In a more advanced approach,
one can use a $G$ matrix obtained self-consistently from
the $s$-channel exchange versions of the Dyson equation shown in Fig.~\ref{1} and the Eliashberg equations shown in Fig.~\ref{fig20}.
The $s$-channel exchange makes the situation look differently from that in the $ t $-channel exchange.
In OBE models, a limited set of diagrams can be evaluated.
In QCB models there are no loop corrections to the $ d_{\alpha} NN $ vertex,
so that all diagrams can be summed up. It is expected therefore that QCB models are exactly (numerically) solvable.
This should not seem surprising, given the analytically solvable Lee model is the predecessor of QCB models.

We thus reproduced successfully the pairing gap of superfluid neutrons below the saturation density.
The results demonstrate great potential of the $s$-channel exchange interaction models in a realistic description of nuclear matter.

\begin{acknowledgments}
This work was supported in part by RFBR Grant No.~16-02-01104 and Grant~No.~HLP-2015-18~of Heisenberg-Landau~Program.

%%%%%%%%%%%%%%%%%%%%%%%%%

%%%%%%%%%%%%%%%%%%%%%%%%%

\end{acknowledgments}


\begin{thebibliography}{0}
\bibitem{4} Using the Holtzmark distribution, the average distance from the nucleon to its nearest neighbor nucleon at the saturation density is estimated as $\langle r \rangle = 1.02$ fm with a statistical error of $\pm$ 0.37 fm \cite{14}. This distance seems surprisingly small in comparison with both the proton charge radius $\langle r^2_{p}\rangle^{1/2} = 0.875 \pm 0.007$ fm, and the pion charge radius $\langle r^2_{\pi} \rangle^{1/2} = 0.659 \pm 0.025$ fm \cite{5}. A significant overlap of nucleons in nuclei practically does not leave space for mesons to participate in the exchange processes.
\bibitem{26}
As can be seen from Fig.~\ref{fig2} of Ref.~\cite{25},
nuclear matter with virtual dibaryons below the critical density of a dibaryon Bose condensation
is more stiff
than nuclear matter without dibaryon degrees of freedom.
\end{thebibliography}
\end{document}